\documentclass[sigconf]{acmart}

\AtBeginDocument{%
  \providecommand\BibTeX{{%
    \normalfont B\kern-0.5em{\scshape i\kern-0.25em b}\kern-0.8em\TeX}}}

\setcopyright{acmcopyright}
\copyrightyear{2018}
\acmYear{2018}
\acmDOI{XXXXXXX.XXXXXXX}

\acmConference[Conference acronym 'XX]{Make sure to enter the correct
  conference title from your rights confirmation emai}{June 03--05,
  2018}{Woodstock, NY}
\acmPrice{15.00}
\acmISBN{978-1-4503-XXXX-X/18/06}

\begin{document}

\title[FIN: A Novel Spatial-Temporal Modeling Based on Long Sequential Behavior for CTR Prediction]{Fragment and Integrate Network (FIN): A Novel Spatial-Temporal Modeling Based on Long Sequential Behavior for Online Food Ordering Click-Through Rate Prediction}


\author{Jun Li, Jingjian Wang, Hongwei Wang, Xing Deng}
\author{Jielong Chen, Bing Cao, Zekun Wang, Guanjie Xu, Ge Zhang, Feng Shi, Hualei Liu}
\affiliation{%
  \institution{Alibaba Group}
  \city{Beijing}
  \country{China}}
\email{{lj251092,whw383259,jielong.cjl,bingcao.cb,hengkun.wzk,guanjie.xgj,luoge.zg,sam.sf}@alibaba-inc.com,jingjianwang.wjj@koubei.com,dengxing.dx@autonavi.com,jianglang@taobao.com}
\authornote{J. Li and G. Zhang are the corresponding authors. \{lj251092,luoge.zg\}@alibaba-inc.com.}

\renewcommand{\shortauthors}{Jun Li and Ge Zhang, et al.}

\begin{abstract}
Spatial-temporal information has been proven to be of great significance for click-through rate prediction tasks in online Location-Based Services (LBS), especially in mainstream food ordering platforms such as DoorDash, Uber Eats, Meituan, and Ele.me. Modeling user spatial-temporal preferences with sequential behavior data has become a hot topic in recommendation systems and online advertising. However, most of existing methods either lack the representation of rich spatial-temporal information or only handle user behaviors with limited length, e.g. 100. In this paper, we tackle these problems by designing a new spatial-temporal modeling paradigm named \textbf{F}ragment and \textbf{I}ntegrate \textbf{N}etwork (FIN). FIN consists of two networks: (i) Fragment Network (FN) extracts \textbf{M}ultiple \textbf{S}ub-\textbf{S}equences (MSS) from lifelong sequential behavior data, and captures the specific spatial-temporal representation by modeling each MSS respectively. Here both a simplified attention and a complicated attention are adopted to balance the performance gain and resource consumption. (ii) Integrate Network (IN) builds a new integrated sequence by utilizing spatial-temporal interaction on MSS and captures the comprehensive spatial-temporal representation by modeling the integrated sequence with a complicated attention. Both public datasets and production datasets have demonstrated the accuracy and scalability of FIN. Since 2022, FIN has been fully deployed in the recommendation advertising system of Ele.me, one of the most popular online food ordering platforms in China, obtaining 5.7\% improvement on Click-Through Rate (CTR) and 7.3\% increase on Revenue Per Mille (RPM).
\end{abstract}





\maketitle

\section{Introduction}
Online Food Ordering Service (OFOS) is a popular location-based service that helps people order the food they want. Unlike traditional e-commerce scenarios, the spatial and temporal information has a great impact on users' behavioral preferences in OFOS platforms such as DoorDash, Uber Eats, Meituan, and Ele.me\footnote{\url{https://www.ele.me/}}. For example, the probability of a user ordering at a restaurant decreases as the spatial distance between the restaurant and the user increases. Additionally, the categories of food ordered by users vary significantly across different periods of the day. Specifically, people tend to order fast food during lunchtime at their workplace, while they may prefer kebab in the evening at his residence. A user's dietary preferences are reflected in their historical behavior data \cite{lin2022spatiotemporalenhanced}. Over 22\% of users in Ele.me, have more than 1000 click behaviors in the last 12 months. User behavior sequence data has a lot of spatial-temporal information, which makes it ideal for learning user preference in different spatial-temporal contexts.

Recently, modeling spatial-temporal information with sequential behavior has become a hot research topic \cite{wang2020calendar,DBLP:journals/corr/abs-2006-05639,lin2022spatiotemporalenhanced}. StEN \cite{lin2022spatiotemporalenhanced} activates relevant preferences from user behavior sequences using Feed Forward Network (FFN) and average-pooling with time periods and location information.
BASM \cite{du2022basm} filters behavior sequences with time periods and locations, and then generates dynamic network parameters through a meta-network. However, these approaches can only handle user behaviors with limited length and lack the representation of comprehensive spatial-temporal information. SIM \cite{DBLP:journals/corr/abs-2006-05639}, UBR4CTR \cite{Qin_2020_UBR4CTR}, and ETA \cite{DBLP:journals/corr/abs-2108-04468} adopt two cascading units, one of which extracts the top-K relevant behaviors as a sub-sequence from long behavior sequence, while the other adopts the complicated attention to model the precise relationship between query items and the sub-sequence. However, these approaches can not involve rich spatial-temporal information of long sequential behavior data, which is of great value for ranking in online food ordering scenarios.

In this paper, we address the above problems by designing a new spatial-temporal modeling paradigm named as \textbf{F}ragment and \textbf{I}ntegrate \textbf{N}etwork (FIN). Inspired by the ideas of SIM \cite{DBLP:journals/corr/abs-2006-05639} and StEN \cite{lin2022spatiotemporalenhanced}, FIN can capture complex spatial-temporal intent representation of lifelong behavior sequences. FIN consists of a Fragment Network (FN) and an Integrate Network (IN). FN captures the specific spatial-temporal representation by modeling each MSS respectively. IN captures the comprehensive spatial-temporal representation by modeling a integrated sequence.

The main contributions of this paper are summarized as follows:

\begin{itemize}
    \item We propose a new modeling paradigm FIN for modeling spatial-temporal intent representation based on long behavior sequences, of which FN can model the specific spatial-temporal representation in MSS, while IN can model the comprehensive spatial-temporal representation among MSS.
\end{itemize}

\begin{itemize}
    \item In FN, a simplified attention is proposed to balance the performance gain and resource consumption for modeling long sub-sequence. In IN, a new spatial-temporal interaction is proposed to capture the comprehensive spatial-temporal representation among MSS.
\end{itemize}

\begin{itemize}
    \item Since 2022, FIN has been fully deployed in the recommendation advertising system in Ele.me, one of the most popular online food ordering platforms in China, bringing 5.7\% CTR and 7.3\% RPM lift.
\end{itemize}

\begin{itemize}
    \item The design of Fragment and Integrate mechanism enables FIN with a better ability to deploy in both scalability and accuracy, providing a new approach for modeling lifelong sequential behavior data in recommendation systems and online advertising.
\end{itemize}

\section{Related Work}
\textbf{Long-term User Interest.} Deep learning based methods have achieved great success in CTR prediction task \cite{DBLP:journals/corr/ChengKHSCAACCIA16, wang2017deep, Lian_2018, guo2017deepfm}. Recently, a series of works \cite{Wang_srs2019, feng2019dsin, bian2021can} on modeling user behavior sequence have emerged. DIN \cite{zhou2018din}, DIEN \cite{zhou2018dien}, MIND \cite{li2019multiinterest} and Transformer \cite{DBLP:journals/corr/VaswaniSPUJGKP17} usually model short-term user behaviors due to the limitation of
latency. MIMN \cite{DBLP:journals/corr/abs-1905-09248} has proven that long sequence behaviors can significantly improve the performance of CTR models, adopting a memory network to compress long user behavior into a fixed-sized interest memory. SIM \cite{DBLP:journals/corr/abs-2006-05639} selects top-K behaviors as a sub-sequence from lifelong sequence, and models the sub-sequence using multi-head attention. In a similar way, ETA \cite{DBLP:journals/corr/abs-2108-04468}, SDIM \cite{cao2022sampling}, and TWIN \cite{chang2023twin} propose an end-to-end method to address the inconsistency between GSU and ESU of SIM \cite{DBLP:journals/corr/abs-2006-05639}. In the field of NLP and CV, FLASH \cite{hua2022transformer} and GSA-CCA \cite{jung2022grouped} utilize both local attention and global attention to model long sequences. However, these works mentioned above rarely involve modeling the spatial-temporal information in long behavior sequence.

\textbf{Spatial-temporal Modeling.} The CTR prediction task can be represented as the likelihood of a user clicking on an item in a certain 
context, of which the spatial-temporal information has been proven to be of great value \cite{Ouyang_2019,inproceedings_Agarwal09, 10.1145/3511808.3557458}. Recently, a series of works has emerged combining spatial-temporal information with behavior sequences \cite{DBLP:journals/corr/abs-2006-05639,Yang23_new_extraction,lin2022spatiotemporalenhanced}. SLi-Rec \cite{Yu_Adaptive19} models user long-term and short-term interests by introducing time-aware and content-aware controllers. TRISAN \cite{Qi_Trilateral21} leverages a triangular relationship between user geographic location, item geographic location, and user click time to enhance the representation of spatial-temporal information. BASM \cite{du2022basm} proposes a bottom-up network to model multiple spatial-temporal data distributions. StEN \cite{lin2022spatiotemporalenhanced} proposes a spatial-temporal enhancement network to model the spatial-temporal representation of user behavior data. However, these models can only handle user behaviors with limited length and lack the representation of comprehensive spatial-temporal relationships in long sequential behavior data. 
\begin{figure*}
  \centering
  \includegraphics[trim=0 0 0 20, width=\linewidth]{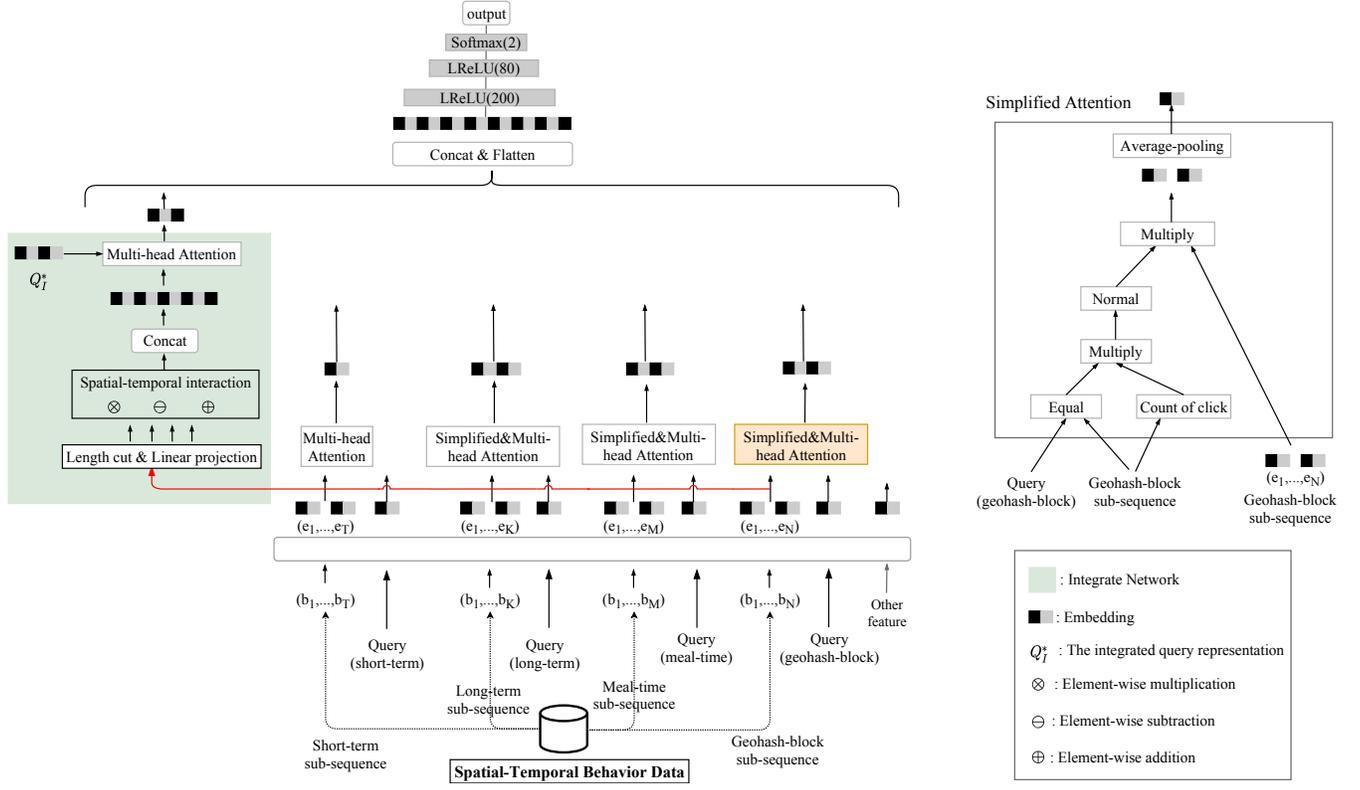}
  \caption{Fragment and Integrate Network (FIN). It follows the traditional Embedding\&MLP paradigm which takes the output of the Fragment Network, the Integrate Network and other features as inputs.}
\end{figure*}
\section{Fragment And Integrate Network}
Both the spatial-temporal information and sequential behavior data have been proved to be effective for CTR prediction tasks. BASM \cite{du2022basm} and StEN \cite{lin2022spatiotemporalenhanced} adopt an adaptive spatial-temporal network to model user behavior sequences with limited length. SIM \cite{DBLP:journals/corr/abs-2006-05639} and ETA \cite{DBLP:journals/corr/abs-2108-04468} model long sequential behavior data with limited temporal information (time intervals). How to efficiently model the rich spatial-temporal information based on long behavior sequences has become a major challenge, particularly for online LBS.

To tackle this challenge, we propose a new modeling named\textbf{ F}ragment and\textbf{ I}ntegrate \textbf{N}etwork (FIN), which can both efficiently model long behavior sequences and effectively learn the rich spatial-temporal information. In this section, we will first introduce overall architecture of FIN, and then introduce the two proposed networks in detail.

\subsection{Overview Architecture}
The overall architecture of FIN is shown in Figure 1, consisting of a Fragment Network (FN) and an Integrate Network (IN).

\textbf{In FN}, four sub-sequences are extracted, following a strategy of spatial-temporal search. A simplified attention is proposed to model the long sub-sequences, followed by a complicated multi-head attention to model the truncated dozens of behaviors in these sub-sequences respectively.

\textbf{In IN}, we build a new integrated sequence from the four sub-sequences by utilizing spatial-temporal interactions so that the comprehensive spatial-temporal representation can be learned with a complicated multi-head attention. It is worth mentioning that the spatial-temporal interaction has a practical physical meaning and is compatible among multiple sub-sequences with various lengths and dimensions.

\subsection{Fragment Network}

It is impractical to directly model all user behavior sequences in large-scale recommendation systems and online advertising, due to limited resource and response time. Inspired by category-based search proposed in SIM \cite{DBLP:journals/corr/abs-2006-05639}, we propose a hard search method based on spatial-temporal information: 

\begin{itemize}
    \item \textbf{Geohash-block}: Divide the latitude and longitude data into a finite geohash\footnote{\url{https://en.wikipedia.org/wiki/Geohash}} code block. Use the geohash block as a query to search the user's lifelong behavior sequence and extract a Geohash-block sub-sequence.
\end{itemize}

\begin{itemize}
    \item \textbf{Meal-time}: Divide 24 hours of a day into several meal-time periods by bucketing minutes in our sample and extract a Meal-time sub-sequence using the periods.
\end{itemize}

\begin{itemize}
    \item \textbf{Short-term}: Select a period as Short-term, e.g. the recent 30 days. Use it to extract a short-term sub-sequence.
\end{itemize}

\begin{itemize}
    \item \textbf{Long-term}: Select another period as Long-term, e.g. the recent 365 days and build the long-term sub-sequence in a similar way to Geohash-block.
\end{itemize}

Here we introduce methods for modeling these sub-sequences respectively: 

\textbf{Geohash-block Modeling}: We convert each latitude-longitude pair $(l_1, l_2)$ into a 6-digit alphanumeric string \textbf{geohash6}. Given the list of long user behaviors $B=[b_1;b_2;...;b_i;...;b_N]$, where $b_i$ is the $i$-th user behavior and N is the length of user behavior list. Then we utilize geohash6 as a query to retrieve a sub-sequence $B^*$ under the same geohash6 from $B$. 

Because the length of $B^*$ is still long, up to several hundreds, we firstly adopt a simplified attention to model it. Given the \textbf{S}ide information of \textbf{Q}uery item\footnote{In this paper, query item refers to the item aimed to be scored by the CTR model.} $SQ=[sq_1;sq_2;...;sq_i;...;sq_K]$, where $sq_i$ is $i$-th side information (such as category-id, item-id, geohash, meal-time, etc), and $K$ is the number of side information. Let $SB^i=[sb_1^i;sb_2^i;...;sb_j^i;...;sb_T^i]$ be $i$-th \textbf{S}ide information in $B^*$, where $sb_j^i$ is the $j$-th user behavior with $i$-th side information, and $T$ is the length of $B^*$. The $SB^i$ is encoded as embedding $E^i=[e_1^i;e_2^i;...;e_j^i;...;e_T^i]$. We equal $sb_j^i$ with $sq_i$ and generate relevant score $r_j^i$:
\begin{equation}
  r_j^i=sign(sb_j^i=sq_i)
\end{equation}
Let $C=[c_1;c_2;...;c_j;...;c_T]$ be the count of clicks in $B^*$, where $c_j$ is the cumulative clicks for $j$-th item in $B^*$. Note that $c_j$ may be greater than 1, since we de-duplicate the items in $B^*$. The normalization score $z_{score}^i$ of $i$-th side information is obtained by multiplying the $r_j^i$ and $c_j$, and divided by total clicks $c_j$:
\begin{equation}
  z_{score}^i=\frac{\sum_{j=1}^{T}r_j^i c_j}{\sum_{j=1}^{T}c_j}
\end{equation}
The simplified attention representation $U^*$ is obtained by multiplying $z_{score}^i$ and $e_j^i$ with pooling and concatenate the $K$ representations $U_K$:
\begin{equation}
  U_i=pooling(z_{score}^i e_j^i)
\end{equation}
\begin{equation}
  U^*=concat(U_1;...;U_K)
\end{equation}

To model $B^*$ precisely, we select $B_c^*$ by truncating the length of $B^*$ to the latest dozens and take advantage of the complicated multi-head attention to model $B_c^*$ with different query items: 
\begin{equation}
  att_m=softmax(W_{km}e_{k} \odot W_{qm}e_{q})
\end{equation}
\begin{equation}
  head_m=att_m \odot W_{vm}e_{k}
\end{equation}
\begin{equation}
  U_{c}^*=concat(head_1;...;head_N)
\end{equation}
where $e_{k}$ and $e_{q}$ denote the embedding of $B_c^*$ and query item respectively. $W_{km}$, $W_{qm}$, $W_{vm}$ are the $m$-th matrix parameter of $head_m$. $att_m$ is the $m$-th attention score, and $head_m$ is $m$-th head in multi-head attention. The $N$ heads are concatenated as the representation $U_{c}^*$.

Finally, we concatenate $U^*$ with $U_{c}^*$ as the output of Geohash-block sub-sequence modeling.

\textbf{Meal-time Modeling}: According to dietary habits\footnote{\url{https://en.wikipedia.org/wiki/Outline_of_meals}}, meal-time can be roughly divided into five periods: breakfast, lunch, afternoon tea, dinner, and late-night snack. Here, based on the distribution of our industrial samples, we divide the meal-time into some fine-grained time periods $P=[p_1;...;p_M]$ by bucketing the minute-level data $T$ using equal-frequency binning \cite{Kotsiantis2006DiscretizationTA}, here $M$ as 95. Given the current time $HH:MM:SS$, $T$ and $P$ can be formulated as:
\begin{equation}
  T=HH*60+MM
\end{equation}
\begin{equation}
  P=quantile(T)
\end{equation}
We use the query $P$ to retrieve a sub-sequence $B^*$ under the same $P$ from $B$. Then the simplified attention and multi-head attention are adopted respectively, in a similar way to the formulations shown in Eq. (1-7).

Additionally, practical experience indicates that user behaviors in $B$, include not only static features, such as item-id, category-id, brand-id, geohash, etc., but also contextual features such as time intervals, stay time, weekdays, weekends, delivery distance, furthermore, they also include statistical features such as impressions and clicks in recent days. These pieces of information are particularly important for online food ordering CTR prediction.

\textbf{Short-term Modeling}: Recent user behaviors, including real-time actions, are crucial for the user's next decision. To capture short-term patterns, the latest dozens of behaviors are extracted from the user's recent 30-day sequence to form a sub-sequence $B^*$. Then we model $B^*$ using multi-head attention.

\textbf{Long-term de-duplicate Modeling}: Some user behaviors $B^\#$ are not considered by Geohash-block, Meal-time, and Short-term. The $B^\#$ is mainly composed of long-term user behavior in non-current Geohash-block or non-current Meal-time. In Ele.me, in the last 6 months, over 38\% of users have geohash6 with a number of more than 3, while over 57\% of users have meal-time with a number of more than 3. Capturing $B^\#$ is important for the comprehensive understanding of user preferences. To this end, we initially borrow ETA \cite{DBLP:journals/corr/abs-2108-04468} and model the whole user behaviors with a length of 1024 in last 12 months, but the AUC score does not show a significant improvement. In online food ordering scenarios, people tends to order meals in the same restaurant repeatedly, since the region in which users can order meals is limited with geographic constraints \cite{lin2022spatiotemporalenhanced}. Therefore, we de-duplicate the restaurants in the behavior sequence, and add behavior frequency and time interval to the side information of the behavior sequence. The de-duplicate sequence length is reduced by 3.5 times. Our experiments found that the performance of modeling the de-duplicate sequence truncated to latest 100 using multi-head attention, is superior to ETA which selects top-64 from a long sequence of length 500. Given constraints of storage and computing cost, we adopt multi-head attention to model the truncated long-term sub-sequence.

It is worth noting that the modeling of these sub-sequences mentioned above is parallel and independent of each other, which ensures deploying flexibility and serving efficiency. The output of each sub-sequence modeling represents a specific user intent representation in a specific spatial-temporal context, respectively.

\subsection{Integrate Network}

We can learn the comprehensive representation of spatial-temporal interest by building a new integrated sequence using spatial-temporal interactions on these sub-sequences of Meal-time, Geohash-block, Long-term, and Short-term. The spatial-temporal interaction has a practical physical meaning, as shown in Figure 2.
\begin{figure}[h]
  \centering
  \includegraphics[trim=0 20 0 20, width=\linewidth]{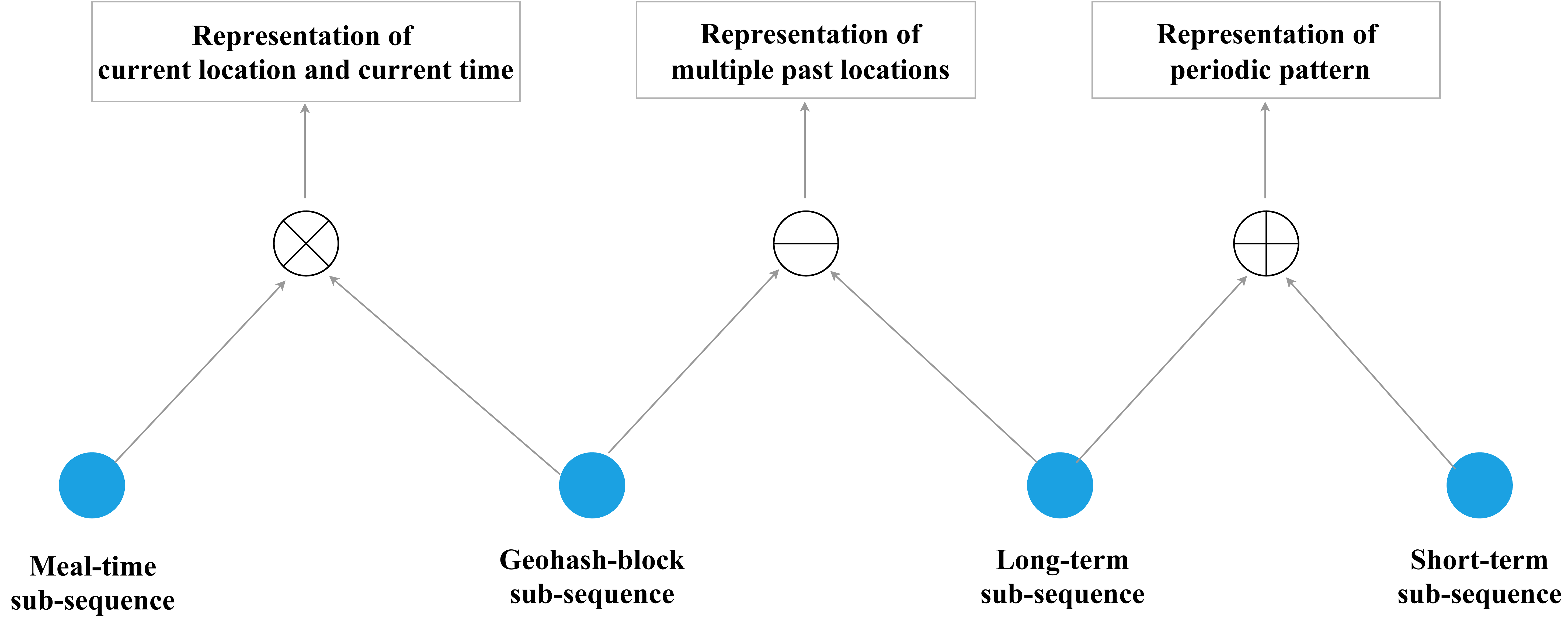}
  \caption{The cases that explain the practical physical meaning of spatial-temporal interaction on multiple sub-sequences.}
\end{figure}

When implementing the integrate network, we encounter a practical issue: both the lengths and dimensions of each sub-sequence are different. A naive solution is to average the lengths of sub-sequences to 1 and align the dimensions of sub-sequences through linear projection. Although this method can improve AUC to some extent, averaging the lengths to 1 may result in the loss of behavior information, we ultimately adopt a spatial-temporal interaction based on the granularity of sequence elements instead. Let $B_G$, $B_M$, $B_S$ and $B_L$ be the sub-sequences of Geohash-block, Meal-time, Short-term and Long-term, respectively. $B_G^*$, $B_M^*$, $B_S^*$ and $B_L^*$ are obtained by truncating the lengths of the sub-sequences to dozens, and aligning the dimensions through linear projection \cite{DBLP:journals/corr/VaswaniSPUJGKP17}. The spatial-temporal
interaction can be formulated as:
\begin{equation}
  B_i^*= cross(B_G^*, B_M^*, B_S^*, B_L^*)
\end{equation}
\begin{equation}
  U_{I}^*=concat(B_1^*;...;B_i^*;...;B_K^*)
\end{equation}
where $cross$ is the set of operations 
including element-wise multiplication, subtraction, and addition. The $U_{I}^*$ is the integrated user representation by concatenating $B_i^*$. Let $Q_G$, $Q_M$, $Q_S$ and $Q_L$ be the query with respect to these sub-sequences, we can also obtain the integrated query representation $Q_{I}^*$, in a similar way to $U_{I}^*$. Then we take advantage of multi-head attention to capture the comprehensive spatial-temporal interest representation between $U_{I}^*$ and $Q_{I}^*$ by using the formulations shown in Eq. (5-7).

Both the output of Fragment Network and Integrate Network are concatenated as the final representation of user spatial-temporal interest for CTR prediction.

\section{Experiments}

In this section, we present our experiments in detail, including the datasets, compared models, implementation details, and some corresponding analyses. The proposed FIN is compared with several state-of-the-art works on two public datasets and one industrial dataset as shown in Table 1. To demonstrate the practicality and scalability of the proposed method, we utilize category, price buckets, and geohash information to represent the spatial-temporal information on public datasets. We also conduct careful online A/B testing, with a comparison of several SOTA industry models.

\subsection{Experimental Settings}
\textbf{Datasets.}
The statistical metrics of all datasets is shown in Table 1. 

\begin{table}[h]
  \caption{The statistical data of the datasets used in this paper}
  \label{tab:freq}
  \footnotesize
  \begin{tabular}{lcccc}
    \toprule
    Dataset&Users&Items&Categories&Instances\\
    \midrule
    Amazon.&98613&130880&1400&197226\\
    Google Local.&286588&141909&2810&584176\\
    Industrial.&70.3 million&2.8 million &173&10.7 billion\\
  \bottomrule
\end{tabular}
\end{table}

\textbf{Amazon dataset}\footnote{\url{https://cseweb.ucsd.edu/~jmcauley/datasets/amazon/links.html}} consists of product reviews and metadata from Amazon \cite{He_2016}. We use the Books subset, and choose price data equifrequency binned with a total of 48 bins and category data to represent the spatial-temporal information. We create corresponding user behavior sequences, and randomly select 80\% of the samples as the training set and 20\% of the samples as the test set, following previous works \cite{DBLP:journals/corr/abs-2006-05639, DBLP:journals/corr/abs-1905-09248}.

\textbf{Google Local dataset}\footnote{\url{https://cseweb.ucsd.edu/~jmcauley/datasets.html\#google_local}} consists of review information and business metadata from Google Maps \cite{li2022uctopic}. We use the New York subset, and encode the latitude and longitude data into 5-digit alphanumeric string (geohash5) with a total of 4223 geohash5. Then Geohash5 and category data are chosen as the spatial-temporal information. The following dataset preprocessing steps are the same as the Amazon dataset.

\textbf{Industrial dataset} is collected from the Ele.me online recommendation advertising system. Samples are constructed from impression logs, with “click” or “not” as the label. Two weeks' samples are used for training, while the samples on the following day for testing. Comparing with the e-commerce scenario, we have fewer categories but richer spatial-temporal information. We convert each latitude and longitude data into a 6-digit alphanumeric string (geohash6) with a total of 1.3 million geohash6, and bucket the minute-level data into meal-time periods with a total of 95 time periods. More than 55\% of the samples contain user behaviors with lengths exceeding 500.

\textbf{Compared models.}
We compare our proposed FIN network with mainstream CTR prediction models, including:

\begin{itemize}
    \item \textbf{DIN} \cite{zhou2018din} is a method that models short-term user behavior using attention.
\end{itemize}

\begin{itemize}
    \item \textbf{Avg-Pooling Long DIN} applies average-pooling on long-term behavior and concatenates with DIN output.
\end{itemize}

\begin{itemize}
    \item \textbf{SIM} \cite{DBLP:journals/corr/abs-2006-05639} We choose SIM (hard) instead of SIM (soft) considering their practicality. The embedding of time intervals is added in all models compared on our industrial dataset.
\end{itemize}

\begin{itemize}
    \item \textbf{ETA} \cite{DBLP:journals/corr/abs-2108-04468} is an end-to-end long sequence modeling method based on the LSH method.
\end{itemize}

\begin{itemize}
    \item \textbf{StEN} \cite{lin2022spatiotemporalenhanced} is a spatial-temporal model based on user behavior with time periods and location information.
\end{itemize}

\begin{itemize}
    \item \textbf{FIN} models the specific spatial-temporal representation in FN and the comprehensive spatial-temporal representation in IN.
\end{itemize}

\textbf{Implementation details.} All models are trained with Adam \cite{kingma2017adam}, the learning rate is set to 0.001. We model the behavior sequence using multi-head attention, the number of heads is set to 4. Layers of fully connected network (FCN) are set by 200 $\times$ 80 $\times$ 2, same as previous works \cite{DBLP:journals/corr/abs-2006-05639, DBLP:journals/corr/abs-1905-09248}. The number of embedding dimension is set to 4. We choose widely-used AUC as model performance measurement metrics.

\subsection{Results on Public Datasets}

Table 2 shows the results of all the compared models. Compared with DIN, Avg-Pooling Long DIN performs better, demonstrating that long-term user behavior is helpful. SIM and ETA outperform Avg-Pooling Long DIN, proving that using multi-head attention is more effective to model user behavior sequences. Compared with SIM and ETA, StEN performs better, indicating that the performance of the two sub-sequences extracted by spatial-temporal information is superior to single sub-sequence modeled exactly. Compared with StEN, FIN achieves significant performance improvement, because StEN drops part of sequence information and tends to ignore the spatial-temporal interaction between multiple sub-sequences. The experimental results have demonstrated that FIN outperforms all other models.

\begin{table}[h]
  \caption{The model performance AUC on the public dataset}
  \label{tab:freq}
  \footnotesize
  \begin{tabular}{lcc}
    \toprule
    Model&Amazon(mean$\pm$std)&Google Local(mean$\pm$std)\\
    \midrule
    DIN&0.7899$\pm$0.00019&0.8945$\pm$0.00044\\
    Avg-Pooling Long DIN&0.7953$\pm$0.00017&0.8988$\pm$0.00036\\
    SIM&0.7975$\pm$0.00014&0.9001$\pm$0.00038\\
    ETA&0.7976$\pm$0.00035&0.9007$\pm$0.00021\\
    StEN&0.7987$\pm$0.00019&0.9108$\pm$0.00031\\
    FIN&\textbf{0.8031$\pm$0.00032}&\textbf{0.9136$\pm$0.00018}\\
  \bottomrule
\end{tabular}
\end{table}

\textbf{Ablation study.} We evaluate the effectiveness of the proposed FIN network by applying different operators to long sequential behavior. As shown in Table 3, both StEN and our proposed Simplified attention outperform SIM. The performance of FN is better than StEN, demonstrating the effectiveness of long-term de-duplicate sub-sequence in FN. Finally, the proposed FIN achieves further performance improvement than FN, demonstrating that the spatial-temporal interaction between multiple sub-sequences can bring rich information gain.

\begin{table}[h]
  \caption{Effectiveness evaluation of the FIN structure}
  \label{tab:freq}
  \footnotesize
  \begin{tabular}{lcc}
    \toprule
    Operations&Amazon(mean$\pm$std)&Google Local(mean$\pm$std)\\
    \midrule
    Avg-Pooling Long DIN&0.7953$\pm$0.00017&0.8988$\pm$0.00036\\
    Simplified attention&0.7977$\pm$0.00017&0.9102$\pm$0.00011\\
    SIM&0.7975$\pm$0.00014&0.9001$\pm$0.00038\\
    StEN&0.7987$\pm$0.00019&0.9108$\pm$0.00031\\
    FN&\textbf{0.8005$\pm$0.00015}&\textbf{0.9124$\pm$0.00029}\\
    FIN&\textbf{0.8031$\pm$0.00032}&\textbf{0.9136$\pm$0.00018}\\
  \bottomrule
\end{tabular}
\end{table}

\subsection{Results on Industrial Dataset}

We further conduct experiments on the dataset collected from the online recommendation advertising system of Ele.me. All models first take Short-term sub-sequence as one of inputs. SIM is the model based on Geohash-block sub-sequence. The K value is set to 64 in ETA. StEN models the Geohash-block sub-sequence and Meal-time sub-sequence. Table 4 shows the results, StEN outperforms SIM and ETA, demonstrating the importance of spatial-temporal information in our online food ordering scenario. FIN performs better than StEN, proving that our spatial-temporal modeling with the fragment and integrate network is significantly more effective to capture users' interest on various restaurants. Compared with SIM, FIN achieves an AUC increase of 0.0066, which is significant for our business.

\begin{table}[h]
  \caption{Model performance (AUC) on industrial datasets}
  \label{tab:freq}
  \begin{tabular}{lc}
    \toprule
    Model&AUC\\
    \midrule
    SIM&0.6786\\
    ETA&0.6798\\
    StEN&0.6823\\
    FN&0.6839\\
    FIN&0.6852\\
  \bottomrule
\end{tabular}
\end{table}

\textbf{Online A/B Testing.} Since 2022, we have deployed our proposed FIN in the recommendation advertising system of Ele.me. From June 17, 2022 to August 1, 2022, we conduct a rigorous online A/B testing experiment to validate the proposed model. Compared to SIM (our last product base model), FIN achieves significantly higher gain in terms of CTR and RPM in the recommendation advertising scene, which shows in Table 5. Since August 2022, FIN  serves the main scene traffic every day, which contributes significant business revenue growth.

\begin{table}[h]
  \caption{Compared to SIM, the improvement rate of FIN's online results in the homepage and category page of Ele.me}
  \label{tab:freq}
  \footnotesize
  \begin{tabular}{lcc}
    \toprule
    Metric&CTR&RPM\\
    \midrule
    Lift rate&5.7\%&7.3\%\\
  \bottomrule
\end{tabular}
\end{table}

\textbf{Case Study.} The proposed FIN performs well in both offline and online evaluations. Will FIN be able to rank items according to the user's current spatial-temporal interest? After the online A/B testing, we analyze the collected click samples from FIN and SIM. FIN introduces temporal interaction information on the basis of spatial information. We analyze the improvement of several typical categories in different time periods, as shown in Figure 3 (the vertical axis values are not clearly indicated, considering commercial data privacy). For the category of milk tea and fruit juice, both the matching efficiency (RPM) and Exposure are improved during the periods of afternoon tea and dinner. For the category of marinated cooked food, the Exposure during the periods of lunch and afternoon tea is greatly improved while ensuring RPM, bringing merchants more orders. For the category of convenience store, the RPM and Exposure during the late night snack period are greatly improved, while the Exposure during the periods of lunch is suppressed. This analysis testifies that FIN does recommend the user relevant items under his current spatial-temporal circumstances.

\begin{figure}[h]
  \centering
  \includegraphics[trim=0 10 0 10, width=\linewidth]{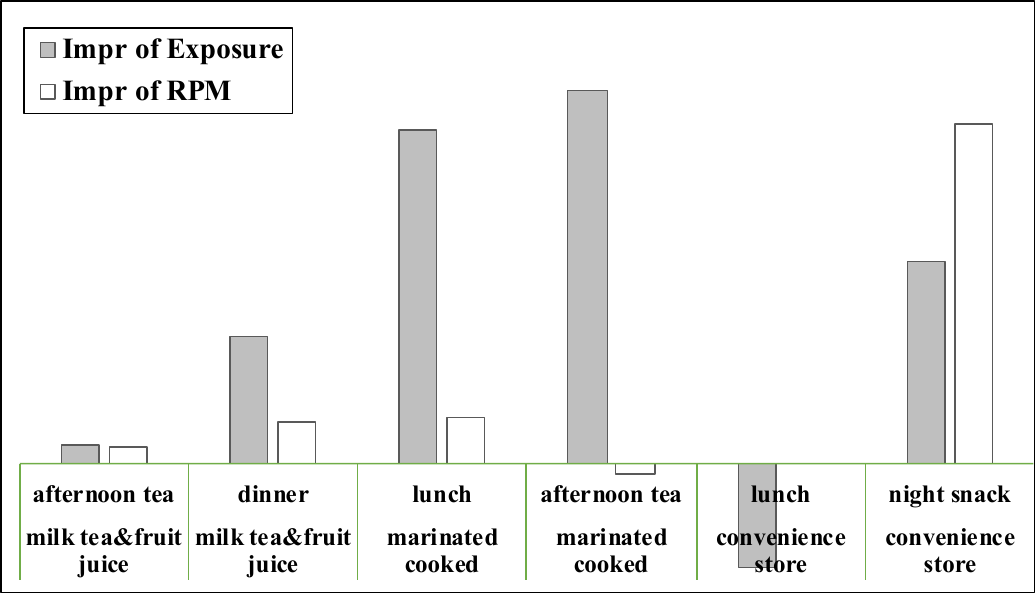}
  \caption{Impr of proportion of Exposure and RPM for typical categories in different time periods from FIN.}
\end{figure}
\textbf{Practical Experience For Deployment.} Here we introduce our practical experience of implementing FIN in our online service system. In Ele.me, the QPS during the lunch peak period can reach up to ten thousand. Real-Time Prediction (RTP) system predicts CTR for hundreds of items in dozens of milliseconds. We decouple real-time behavior data and offline behavior data, so that the real-time behavior data can be updated in seconds, keeping offline behavior data updated daily. The complete behavior sequence is built by combining the real-time behavior data and offline behavior data together in the online service system. The Short-term sub-sequence is consisted of recent dozens of behaviors from the last few days, Geohash-block sub-sequence, Meal-time sub-sequence and Long-term de-duplicate sub-sequence include hundreds of behaviors from the last 12 months, respectively. Furthermore, we optimize the calculation efficiency of multi-head attention in FIN by deep kernel fusion and address the kernel launch overhead issue with CUDA Graph optimization.

\section{CONCLUSIONS}

In this paper, we propose a novel spatial-temporal model based on long sequential behavior data to capture the diverse intent representation of user interest in complex spatial-temporal contexts. We have implemented the proposed FIN in the recommendation advertising system of Ele.me and have brought significant business improvements. We believe that this work will bring new inspiration to spatial-temporal modeling and long behavior sequence modeling. In the future, we will continue to develop spatial-temporal models for various types of user behaviors, such as search behaviors and exposure-unclicked behaviors in the diverse spatial-temporal contexts.

\begin{acks}
The authors would like to thank Jingjing Duan, Xufeng He, Dawei Wang, Lingbao Ye, Juanjuan Li, Weijie Zhuang, Jiajia Shi, Jianghua Ye who did effective optimization work for online deployment.
\end{acks}

\bibliographystyle{ACM-Reference-Format}
\balance
\bibliography{fin}


\begin{thebibliography}{32}


\ifx \showCODEN    \undefined \def \showCODEN     #1{\unskip}     \fi
\ifx \showDOI      \undefined \def \showDOI       #1{#1}\fi
\ifx \showISBNx    \undefined \def \showISBNx     #1{\unskip}     \fi
\ifx \showISBNxiii \undefined \def \showISBNxiii  #1{\unskip}     \fi
\ifx \showISSN     \undefined \def \showISSN      #1{\unskip}     \fi
\ifx \showLCCN     \undefined \def \showLCCN      #1{\unskip}     \fi
\ifx \shownote     \undefined \def \shownote      #1{#1}          \fi
\ifx \showarticletitle \undefined \def \showarticletitle #1{#1}   \fi
\ifx \showURL      \undefined \def \showURL       {\relax}        \fi
\providecommand\bibfield[2]{#2}
\providecommand\bibinfo[2]{#2}
\providecommand\natexlab[1]{#1}
\providecommand\showeprint[2][]{arXiv:#2}

\bibitem[Agarwal et~al\mbox{.}(2009)]%
        {inproceedings_Agarwal09}
\bibfield{author}{\bibinfo{person}{Deepak Agarwal}, \bibinfo{person}{Bee-Chung
  Chen}, {and} \bibinfo{person}{Pradheep Elango}.}
  \bibinfo{year}{2009}\natexlab{}.
\newblock \showarticletitle{Spatio-temporal models for estimating click-through
  rate}.
\newblock \bibinfo{journal}{\emph{WWW'09 - Proceedings of the 18th
  International World Wide Web Conference}}, \bibinfo{pages}{21--30}.
\newblock
\urldef\tempurl%
\url{https://doi.org/10.1145/1526709.1526713}
\showDOI{\tempurl}


\bibitem[Bian et~al\mbox{.}(2021)]%
        {bian2021can}
\bibfield{author}{\bibinfo{person}{Weijie Bian}, \bibinfo{person}{Kailun Wu},
  \bibinfo{person}{Lejian Ren}, \bibinfo{person}{Qi Pi},
  \bibinfo{person}{Yujing Zhang}, \bibinfo{person}{Can Xiao},
  \bibinfo{person}{Xiang-Rong Sheng}, \bibinfo{person}{Yong-Nan Zhu},
  \bibinfo{person}{Zhangming Chan}, \bibinfo{person}{Na Mou},
  \bibinfo{person}{Xinchen Luo}, \bibinfo{person}{Shiming Xiang},
  \bibinfo{person}{Guorui Zhou}, \bibinfo{person}{Xiaoqiang Zhu}, {and}
  \bibinfo{person}{Hongbo Deng}.} \bibinfo{year}{2021}\natexlab{}.
\newblock \bibinfo{title}{CAN: Feature Co-Action for Click-Through Rate
  Prediction}.
\newblock
\newblock
\showeprint[arxiv]{2011.05625}~[cs.IR]


\bibitem[Cao et~al\mbox{.}(2022)]%
        {cao2022sampling}
\bibfield{author}{\bibinfo{person}{Yue Cao}, \bibinfo{person}{XiaoJiang Zhou},
  \bibinfo{person}{Jiaqi Feng}, \bibinfo{person}{Peihao Huang},
  \bibinfo{person}{Yao Xiao}, \bibinfo{person}{Dayao Chen}, {and}
  \bibinfo{person}{Sheng Chen}.} \bibinfo{year}{2022}\natexlab{}.
\newblock \bibinfo{title}{Sampling Is All You Need on Modeling Long-Term User
  Behaviors for CTR Prediction}.
\newblock
\newblock
\showeprint[arxiv]{2205.10249}~[cs.IR]


\bibitem[Chang et~al\mbox{.}(2023)]%
        {chang2023twin}
\bibfield{author}{\bibinfo{person}{Jianxin Chang}, \bibinfo{person}{Chenbin
  Zhang}, \bibinfo{person}{Zhiyi Fu}, \bibinfo{person}{Xiaoxue Zang},
  \bibinfo{person}{Lin Guan}, \bibinfo{person}{Jing Lu}, \bibinfo{person}{Yiqun
  Hui}, \bibinfo{person}{Dewei Leng}, \bibinfo{person}{Yanan Niu},
  \bibinfo{person}{Yang Song}, {and} \bibinfo{person}{Kun Gai}.}
  \bibinfo{year}{2023}\natexlab{}.
\newblock \bibinfo{title}{TWIN: TWo-stage Interest Network for Lifelong User
  Behavior Modeling in CTR Prediction at Kuaishou}.
\newblock
\newblock
\showeprint[arxiv]{2302.02352}~[cs.IR]


\bibitem[Chen et~al\mbox{.}(2021)]%
        {DBLP:journals/corr/abs-2108-04468}
\bibfield{author}{\bibinfo{person}{Qiwei Chen}, \bibinfo{person}{Changhua Pei},
  \bibinfo{person}{Shanshan Lv}, \bibinfo{person}{Chao Li},
  \bibinfo{person}{Junfeng Ge}, {and} \bibinfo{person}{Wenwu Ou}.}
  \bibinfo{year}{2021}\natexlab{}.
\newblock \showarticletitle{End-to-End User Behavior Retrieval in Click-Through
  RatePrediction Model}.
\newblock \bibinfo{journal}{\emph{CoRR}}  \bibinfo{volume}{abs/2108.04468}
  (\bibinfo{year}{2021}).
\newblock
\showeprint[arXiv]{2108.04468}
\urldef\tempurl%
\url{https://arxiv.org/abs/2108.04468}
\showURL{%
\tempurl}


\bibitem[Cheng et~al\mbox{.}(2016)]%
        {DBLP:journals/corr/ChengKHSCAACCIA16}
\bibfield{author}{\bibinfo{person}{Heng{-}Tze Cheng}, \bibinfo{person}{Levent
  Koc}, \bibinfo{person}{Jeremiah Harmsen}, \bibinfo{person}{Tal Shaked},
  \bibinfo{person}{Tushar Chandra}, \bibinfo{person}{Hrishi Aradhye},
  \bibinfo{person}{Glen Anderson}, \bibinfo{person}{Greg Corrado},
  \bibinfo{person}{Wei Chai}, \bibinfo{person}{Mustafa Ispir},
  \bibinfo{person}{Rohan Anil}, \bibinfo{person}{Zakaria Haque},
  \bibinfo{person}{Lichan Hong}, \bibinfo{person}{Vihan Jain},
  \bibinfo{person}{Xiaobing Liu}, {and} \bibinfo{person}{Hemal Shah}.}
  \bibinfo{year}{2016}\natexlab{}.
\newblock \showarticletitle{Wide {\&} Deep Learning for Recommender Systems}.
\newblock \bibinfo{journal}{\emph{CoRR}}  \bibinfo{volume}{abs/1606.07792}
  (\bibinfo{year}{2016}).
\newblock
\showeprint[arXiv]{1606.07792}
\urldef\tempurl%
\url{http://arxiv.org/abs/1606.07792}
\showURL{%
\tempurl}


\bibitem[Du et~al\mbox{.}(2022)]%
        {du2022basm}
\bibfield{author}{\bibinfo{person}{Boya Du}, \bibinfo{person}{Shaochuan Lin},
  \bibinfo{person}{Jiong Gao}, \bibinfo{person}{Xiyu Ji},
  \bibinfo{person}{Mengya Wang}, \bibinfo{person}{Taotao Zhou},
  \bibinfo{person}{Hengxu He}, \bibinfo{person}{Jia Jia}, {and}
  \bibinfo{person}{Ning Hu}.} \bibinfo{year}{2022}\natexlab{}.
\newblock \bibinfo{title}{BASM: A Bottom-up Adaptive Spatiotemporal Model for
  Online Food Ordering Service}.
\newblock
\newblock
\showeprint[arxiv]{2211.12033}~[cs.LG]


\bibitem[Feng et~al\mbox{.}(2019)]%
        {feng2019dsin}
\bibfield{author}{\bibinfo{person}{Yufei Feng}, \bibinfo{person}{Fuyu Lv},
  \bibinfo{person}{Weichen Shen}, \bibinfo{person}{Menghan Wang},
  \bibinfo{person}{Fei Sun}, \bibinfo{person}{Yu Zhu}, {and}
  \bibinfo{person}{Keping Yang}.} \bibinfo{year}{2019}\natexlab{}.
\newblock \bibinfo{title}{Deep Session Interest Network for Click-Through Rate
  Prediction}.
\newblock
\newblock
\showeprint[arxiv]{1905.06482}~[cs.IR]


\bibitem[Guo et~al\mbox{.}(2017)]%
        {guo2017deepfm}
\bibfield{author}{\bibinfo{person}{Huifeng Guo}, \bibinfo{person}{Ruiming
  Tang}, \bibinfo{person}{Yunming Ye}, \bibinfo{person}{Zhenguo Li}, {and}
  \bibinfo{person}{Xiuqiang He}.} \bibinfo{year}{2017}\natexlab{}.
\newblock \bibinfo{title}{DeepFM: A Factorization-Machine based Neural Network
  for CTR Prediction}.
\newblock
\newblock
\showeprint[arxiv]{1703.04247}~[cs.IR]


\bibitem[He and McAuley(2016)]%
        {He_2016}
\bibfield{author}{\bibinfo{person}{Ruining He} {and} \bibinfo{person}{Julian
  McAuley}.} \bibinfo{year}{2016}\natexlab{}.
\newblock \showarticletitle{Ups and Downs}. In
  \bibinfo{booktitle}{\emph{Proceedings of the 25th International Conference on
  World Wide Web}}. \bibinfo{publisher}{International World Wide Web
  Conferences Steering Committee}.
\newblock
\urldef\tempurl%
\url{https://doi.org/10.1145/2872427.2883037}
\showDOI{\tempurl}


\bibitem[Hua et~al\mbox{.}(2022)]%
        {hua2022transformer}
\bibfield{author}{\bibinfo{person}{Weizhe Hua}, \bibinfo{person}{Zihang Dai},
  \bibinfo{person}{Hanxiao Liu}, {and} \bibinfo{person}{Quoc~V. Le}.}
  \bibinfo{year}{2022}\natexlab{}.
\newblock \bibinfo{title}{Transformer Quality in Linear Time}.
\newblock
\newblock
\showeprint[arxiv]{2202.10447}~[cs.LG]


\bibitem[Jung et~al\mbox{.}(2022)]%
        {jung2022grouped}
\bibfield{author}{\bibinfo{person}{Bumjun Jung}, \bibinfo{person}{Yusuke
  Mukuta}, {and} \bibinfo{person}{Tatsuya Harada}.}
  \bibinfo{year}{2022}\natexlab{}.
\newblock \bibinfo{title}{Grouped self-attention mechanism for a
  memory-efficient Transformer}.
\newblock
\newblock
\showeprint[arxiv]{2210.00440}~[cs.LG]


\bibitem[Kingma and Ba(2017)]%
        {kingma2017adam}
\bibfield{author}{\bibinfo{person}{Diederik~P. Kingma} {and}
  \bibinfo{person}{Jimmy Ba}.} \bibinfo{year}{2017}\natexlab{}.
\newblock \bibinfo{title}{Adam: A Method for Stochastic Optimization}.
\newblock
\newblock
\showeprint[arxiv]{1412.6980}~[cs.LG]


\bibitem[Kotsiantis and Kanellopoulos(2006)]%
        {Kotsiantis2006DiscretizationTA}
\bibfield{author}{\bibinfo{person}{Sotiris~B. Kotsiantis} {and}
  \bibinfo{person}{Dimitris~N. Kanellopoulos}.}
  \bibinfo{year}{2006}\natexlab{}.
\newblock \showarticletitle{Discretization Techniques: A recent survey}.
\newblock


\bibitem[Li et~al\mbox{.}(2019)]%
        {li2019multiinterest}
\bibfield{author}{\bibinfo{person}{Chao Li}, \bibinfo{person}{Zhiyuan Liu},
  \bibinfo{person}{Mengmeng Wu}, \bibinfo{person}{Yuchi Xu},
  \bibinfo{person}{Pipei Huang}, \bibinfo{person}{Huan Zhao},
  \bibinfo{person}{Guoliang Kang}, \bibinfo{person}{Qiwei Chen},
  \bibinfo{person}{Wei Li}, {and} \bibinfo{person}{Dik~Lun Lee}.}
  \bibinfo{year}{2019}\natexlab{}.
\newblock \bibinfo{title}{Multi-Interest Network with Dynamic Routing for
  Recommendation at Tmall}.
\newblock
\newblock
\showeprint[arxiv]{1904.08030}~[cs.IR]


\bibitem[Li et~al\mbox{.}(2022b)]%
        {li2022uctopic}
\bibfield{author}{\bibinfo{person}{Jiacheng Li}, \bibinfo{person}{Jingbo
  Shang}, {and} \bibinfo{person}{Julian McAuley}.}
  \bibinfo{year}{2022}\natexlab{b}.
\newblock \bibinfo{title}{UCTopic: Unsupervised Contrastive Learning for Phrase
  Representations and Topic Mining}.
\newblock
\newblock
\showeprint[arxiv]{2202.13469}~[cs.CL]


\bibitem[Li et~al\mbox{.}(2022a)]%
        {10.1145/3511808.3557458}
\bibfield{author}{\bibinfo{person}{Yinfeng Li}, \bibinfo{person}{Chen Gao},
  \bibinfo{person}{Xiaoyi Du}, \bibinfo{person}{Huazhou Wei},
  \bibinfo{person}{Hengliang Luo}, \bibinfo{person}{Depeng Jin}, {and}
  \bibinfo{person}{Yong Li}.} \bibinfo{year}{2022}\natexlab{a}.
\newblock \showarticletitle{Spatiotemporal-Aware Session-Based Recommendation
  with Graph Neural Networks} \emph{(\bibinfo{series}{CIKM '22})}.
  \bibinfo{publisher}{Association for Computing Machinery},
  \bibinfo{address}{New York, NY, USA}, \bibinfo{pages}{1209–1218}.
\newblock
\showISBNx{9781450392365}
\urldef\tempurl%
\url{https://doi.org/10.1145/3511808.3557458}
\showDOI{\tempurl}


\bibitem[Lian et~al\mbox{.}(2018)]%
        {Lian_2018}
\bibfield{author}{\bibinfo{person}{Jianxun Lian}, \bibinfo{person}{Xiaohuan
  Zhou}, \bibinfo{person}{Fuzheng Zhang}, \bibinfo{person}{Zhongxia Chen},
  \bibinfo{person}{Xing Xie}, {and} \bibinfo{person}{Guangzhong Sun}.}
  \bibinfo{year}{2018}\natexlab{}.
\newblock \showarticletitle{{xDeepFM}}. In
  \bibinfo{booktitle}{\emph{Proceedings of the 24th {ACM} {SIGKDD}
  International Conference on Knowledge Discovery {\&}amp Data Mining}}.
  \bibinfo{publisher}{{ACM}}.
\newblock
\urldef\tempurl%
\url{https://doi.org/10.1145/3219819.3220023}
\showDOI{\tempurl}


\bibitem[Lin et~al\mbox{.}(2022)]%
        {lin2022spatiotemporalenhanced}
\bibfield{author}{\bibinfo{person}{Shaochuan Lin}, \bibinfo{person}{Yicong Yu},
  \bibinfo{person}{Xiyu Ji}, \bibinfo{person}{Taotao Zhou},
  \bibinfo{person}{Hengxu He}, \bibinfo{person}{Zisen Sang},
  \bibinfo{person}{Jia Jia}, \bibinfo{person}{Guodong Cao}, {and}
  \bibinfo{person}{Ning Hu}.} \bibinfo{year}{2022}\natexlab{}.
\newblock \bibinfo{title}{Spatiotemporal-Enhanced Network for Click-Through
  Rate Prediction in Location-based Services}.
\newblock
\newblock
\showeprint[arxiv]{2209.09427}~[cs.IR]


\bibitem[Ouyang et~al\mbox{.}(2019)]%
        {Ouyang_2019}
\bibfield{author}{\bibinfo{person}{Wentao Ouyang}, \bibinfo{person}{Xiuwu
  Zhang}, \bibinfo{person}{Li Li}, \bibinfo{person}{Heng Zou},
  \bibinfo{person}{Xin Xing}, \bibinfo{person}{Zhaojie Liu}, {and}
  \bibinfo{person}{Yanlong Du}.} \bibinfo{year}{2019}\natexlab{}.
\newblock \showarticletitle{Deep Spatio-Temporal Neural Networks for
  Click-Through Rate Prediction}. In \bibinfo{booktitle}{\emph{Proceedings of
  the 25th {ACM} {SIGKDD} International Conference on Knowledge Discovery
  {\&}amp Data Mining}}. \bibinfo{publisher}{{ACM}}.
\newblock
\urldef\tempurl%
\url{https://doi.org/10.1145/3292500.3330655}
\showDOI{\tempurl}


\bibitem[Pi et~al\mbox{.}(2019)]%
        {DBLP:journals/corr/abs-1905-09248}
\bibfield{author}{\bibinfo{person}{Qi Pi}, \bibinfo{person}{Weijie Bian},
  \bibinfo{person}{Guorui Zhou}, \bibinfo{person}{Xiaoqiang Zhu}, {and}
  \bibinfo{person}{Kun Gai}.} \bibinfo{year}{2019}\natexlab{}.
\newblock \showarticletitle{Practice on Long Sequential User Behavior Modeling
  for Click-Through Rate Prediction}.
\newblock \bibinfo{journal}{\emph{CoRR}}  \bibinfo{volume}{abs/1905.09248}
  (\bibinfo{year}{2019}).
\newblock
\showeprint[arXiv]{1905.09248}
\urldef\tempurl%
\url{http://arxiv.org/abs/1905.09248}
\showURL{%
\tempurl}


\bibitem[Pi et~al\mbox{.}(2020)]%
        {DBLP:journals/corr/abs-2006-05639}
\bibfield{author}{\bibinfo{person}{Qi Pi}, \bibinfo{person}{Xiaoqiang Zhu},
  \bibinfo{person}{Guorui Zhou}, \bibinfo{person}{Yujing Zhang},
  \bibinfo{person}{Zhe Wang}, \bibinfo{person}{Lejian Ren},
  \bibinfo{person}{Ying Fan}, {and} \bibinfo{person}{Kun Gai}.}
  \bibinfo{year}{2020}\natexlab{}.
\newblock \showarticletitle{Search-based User Interest Modeling with Lifelong
  Sequential Behavior Data for Click-Through Rate Prediction}.
\newblock \bibinfo{journal}{\emph{CoRR}}  \bibinfo{volume}{abs/2006.05639}
  (\bibinfo{year}{2020}).
\newblock
\showeprint[arXiv]{2006.05639}
\urldef\tempurl%
\url{https://arxiv.org/abs/2006.05639}
\showURL{%
\tempurl}


\bibitem[Qi et~al\mbox{.}(2021)]%
        {Qi_Trilateral21}
\bibfield{author}{\bibinfo{person}{Yi Qi}, \bibinfo{person}{Ke Hu},
  \bibinfo{person}{Bo Zhang}, \bibinfo{person}{Jia Cheng}, {and}
  \bibinfo{person}{Jun Lei}.} \bibinfo{year}{2021}\natexlab{}.
\newblock \showarticletitle{Trilateral Spatiotemporal Attention Network for
  User Behavior Modeling in Location-Based Search}. In
  \bibinfo{booktitle}{\emph{Proceedings of the 30th ACM International
  Conference on Information {\&}amp; Knowledge Management}} (Virtual Event,
  Queensland, Australia) \emph{(\bibinfo{series}{CIKM '21})}.
  \bibinfo{publisher}{Association for Computing Machinery},
  \bibinfo{address}{New York, NY, USA}, \bibinfo{pages}{3373–3377}.
\newblock
\showISBNx{9781450384469}
\urldef\tempurl%
\url{https://doi.org/10.1145/3459637.3482206}
\showDOI{\tempurl}


\bibitem[Qin et~al\mbox{.}(2020)]%
        {Qin_2020_UBR4CTR}
\bibfield{author}{\bibinfo{person}{Jiarui Qin}, \bibinfo{person}{Weinan Zhang},
  \bibinfo{person}{Xin Wu}, \bibinfo{person}{Jiarui Jin},
  \bibinfo{person}{Yuchen Fang}, {and} \bibinfo{person}{Yong Yu}.}
  \bibinfo{year}{2020}\natexlab{}.
\newblock \showarticletitle{User Behavior Retrieval for Click-Through Rate
  Prediction}. In \bibinfo{booktitle}{\emph{Proceedings of the 43rd
  International {ACM} {SIGIR} Conference on Research and Development in
  Information Retrieval}}. \bibinfo{publisher}{{ACM}}.
\newblock
\urldef\tempurl%
\url{https://doi.org/10.1145/3397271.3401440}
\showDOI{\tempurl}


\bibitem[Vaswani et~al\mbox{.}(2017)]%
        {DBLP:journals/corr/VaswaniSPUJGKP17}
\bibfield{author}{\bibinfo{person}{Ashish Vaswani}, \bibinfo{person}{Noam
  Shazeer}, \bibinfo{person}{Niki Parmar}, \bibinfo{person}{Jakob Uszkoreit},
  \bibinfo{person}{Llion Jones}, \bibinfo{person}{Aidan~N. Gomez},
  \bibinfo{person}{Lukasz Kaiser}, {and} \bibinfo{person}{Illia Polosukhin}.}
  \bibinfo{year}{2017}\natexlab{}.
\newblock \showarticletitle{Attention Is All You Need}.
\newblock \bibinfo{journal}{\emph{CoRR}}  \bibinfo{volume}{abs/1706.03762}
  (\bibinfo{year}{2017}).
\newblock
\showeprint[arXiv]{1706.03762}
\urldef\tempurl%
\url{http://arxiv.org/abs/1706.03762}
\showURL{%
\tempurl}


\bibitem[Wang et~al\mbox{.}(2020)]%
        {wang2020calendar}
\bibfield{author}{\bibinfo{person}{Daheng Wang}, \bibinfo{person}{Meng Jiang},
  \bibinfo{person}{Munira Syed}, \bibinfo{person}{Oliver Conway},
  \bibinfo{person}{Vishal Juneja}, \bibinfo{person}{Sriram Subramanian}, {and}
  \bibinfo{person}{Nitesh~V. Chawla}.} \bibinfo{year}{2020}\natexlab{}.
\newblock \bibinfo{title}{Calendar Graph Neural Networks for Modeling Time
  Structures in Spatiotemporal User Behaviors}.
\newblock
\newblock
\showeprint[arxiv]{2006.06820}~[cs.LG]


\bibitem[Wang et~al\mbox{.}(2017)]%
        {wang2017deep}
\bibfield{author}{\bibinfo{person}{Ruoxi Wang}, \bibinfo{person}{Bin Fu},
  \bibinfo{person}{Gang Fu}, {and} \bibinfo{person}{Mingliang Wang}.}
  \bibinfo{year}{2017}\natexlab{}.
\newblock \bibinfo{title}{Deep {\&} Cross Network for Ad Click Predictions}.
\newblock
\newblock
\showeprint[arxiv]{1708.05123}~[cs.LG]


\bibitem[Wang et~al\mbox{.}(2019)]%
        {Wang_srs2019}
\bibfield{author}{\bibinfo{person}{Shoujin Wang}, \bibinfo{person}{Liang Hu},
  \bibinfo{person}{Yan Wang}, \bibinfo{person}{Longbing Cao},
  \bibinfo{person}{Quan~Z. Sheng}, {and} \bibinfo{person}{Mehmet Orgun}.}
  \bibinfo{year}{2019}\natexlab{}.
\newblock \showarticletitle{Sequential Recommender Systems: Challenges,
  Progress and Prospects}. In \bibinfo{booktitle}{\emph{Proceedings of the
  Twenty-Eighth International Joint Conference on Artificial Intelligence}}.
  \bibinfo{publisher}{International Joint Conferences on Artificial
  Intelligence Organization}.
\newblock
\urldef\tempurl%
\url{https://doi.org/10.24963/ijcai.2019/883}
\showDOI{\tempurl}


\bibitem[Yang et~al\mbox{.}(2023)]%
        {Yang23_new_extraction}
\bibfield{author}{\bibinfo{person}{Haifeng Yang}, \bibinfo{person}{Linjing
  Yao}, \bibinfo{person}{Jianghui Cai}, \bibinfo{person}{Yupeng Wang}, {and}
  \bibinfo{person}{Xujun Zhao}.} \bibinfo{year}{2023}\natexlab{}.
\newblock \showarticletitle{A new interest extraction method based on
  multi-head attention mechanism for CTR prediction}.
\newblock \bibinfo{journal}{\emph{Knowledge and Information Systems}}
  (\bibinfo{date}{04} \bibinfo{year}{2023}), \bibinfo{pages}{1--16}.
\newblock
\urldef\tempurl%
\url{https://doi.org/10.1007/s10115-023-01867-w}
\showDOI{\tempurl}


\bibitem[Yu et~al\mbox{.}(2019)]%
        {Yu_Adaptive19}
\bibfield{author}{\bibinfo{person}{Zeping Yu}, \bibinfo{person}{Jianxun Lian},
  \bibinfo{person}{Ahmad Mahmoody}, \bibinfo{person}{Gongshen Liu}, {and}
  \bibinfo{person}{Xing Xie}.} \bibinfo{year}{2019}\natexlab{}.
\newblock \showarticletitle{Adaptive User Modeling with Long and Short-Term
  Preferences for Personalized Recommendation}. In
  \bibinfo{booktitle}{\emph{Proceedings of the 28th International Joint
  Conference on Artificial Intelligence}} (Macao, China)
  \emph{(\bibinfo{series}{IJCAI'19})}. \bibinfo{publisher}{AAAI Press},
  \bibinfo{pages}{4213–4219}.
\newblock
\showISBNx{9780999241141}


\bibitem[Zhou et~al\mbox{.}(2018a)]%
        {zhou2018dien}
\bibfield{author}{\bibinfo{person}{Guorui Zhou}, \bibinfo{person}{Na Mou},
  \bibinfo{person}{Ying Fan}, \bibinfo{person}{Qi Pi}, \bibinfo{person}{Weijie
  Bian}, \bibinfo{person}{Chang Zhou}, \bibinfo{person}{Xiaoqiang Zhu}, {and}
  \bibinfo{person}{Kun Gai}.} \bibinfo{year}{2018}\natexlab{a}.
\newblock \bibinfo{title}{Deep Interest Evolution Network for Click-Through
  Rate Prediction}.
\newblock
\newblock
\showeprint[arxiv]{1809.03672}~[stat.ML]


\bibitem[Zhou et~al\mbox{.}(2018b)]%
        {zhou2018din}
\bibfield{author}{\bibinfo{person}{Guorui Zhou}, \bibinfo{person}{Chengru
  Song}, \bibinfo{person}{Xiaoqiang Zhu}, \bibinfo{person}{Ying Fan},
  \bibinfo{person}{Han Zhu}, \bibinfo{person}{Xiao Ma},
  \bibinfo{person}{Yanghui Yan}, \bibinfo{person}{Junqi Jin},
  \bibinfo{person}{Han Li}, {and} \bibinfo{person}{Kun Gai}.}
  \bibinfo{year}{2018}\natexlab{b}.
\newblock \bibinfo{title}{Deep Interest Network for Click-Through Rate
  Prediction}.
\newblock
\newblock
\showeprint[arxiv]{1706.06978}~[stat.ML]


\end{thebibliography}

\appendix

\end{document}